# Highly tunable, polarization-engineered two-dimensional electron gas in ε-AlGaO$_3$ / ε-Ga$_2$O$_3$ heterostructures

*Praneeth Ranga,[1*], Sung Beom Cho[2,3*], Rohan Mishra[2] and Sriram Krishnamoorthy[1]*

[1] *Department of Electrical and Computer Engineering, The University of Utah, Salt Lake City, UT 84112, United States of America*

[2] *Department of Mechanical Engineering and Materials Science, and Institute of Materials Science and Engineering, Washington University in St. Louis, St. Louis, Missouri 63130, USA*

[3] *Virtual Engineering Center, Korea Institute of Ceramic Engineering and Technology, Jinju-si, Republic of Korea*

**\* indicates equal contribution**

**Email**:praneeth.ranga@utah.edu,rmishra@wustl.edu,sriram.krishnamoorthy@utah.edu

We report on the modeling of polarization-induced two-dimensional electron gas (2DEG) formation at ε-AlGaO$_3$/ε-Ga$_2$O$_3$ heterointerface and the effect of spontaneous polarization (P$_{sp}$) reversal on 2DEG density in ε-Ga$_2$O$_3$/ε-AlGaO$_3$/ε-Ga$_2$O$_3$ double heterostructures. Density-functional theory (DFT) is utilized to calculate the material properties of ε-Ga$_2$O$_3$ and ε-AlGaO$_3$ alloys. Using Schrödinger–Poisson solver along with DFT calculated parameters, the 2DEG density is calculated as a function of barrier type and thickness. By optimizing the layer thicknesses of ε-Ga$_2$O$_3$/ε-AlGaO$_3$/ε-Ga$_2$O$_3$ heterostructures, charge contrast ratios exceeding 1600 are obtained. This computational study indicates the high potential for ε-Ga$_2$O$_3$-based heterostructure devices for non-volatile memories and neuromorphic applications.

Ga$_2$O$_3$ is an emerging ultra-wide bandgap semiconductor with potential applications for high power semiconductor devices and deep-ultraviolet photodetectors[1,2]. Of the five known polymorphs[3], β-Ga$_2$O$_3$ is the most studied phase because of its thermodynamic stability and the availability of high-quality single crystal bulk substrates[4]. β-Ga$_2$O$_3$-based devices have already recorded critical breakdown fields greater than that of GaN and SiC, showing high potential for power device applications[5,6]. High-density two-dimensional electron gases (2DEG) induced by modulation doping of β-(Al$_x$Ga$_{1-x}$)$_2$O$_3$ / β-Ga$_2$O$_3$ heterointerface can potentially lead to increase in 2DEG channel mobility due to reduced ionized-impurity scattering[7–9] and enhanced screening of phonon modes by the 2DEG channel[10]. Attaining high-density 2DEG requires heavy delta-doping in β-(Al$_x$Ga$_{1-x}$)$_2$O$_3$ barrier along with extremely thin spacer layers. This requires tight control of growth parameters and abrupt heterointerfaces along with sharp dopant profiles. Recently, 2DEG sheet charge density of 6.1 x 10$^{12}$ cm$^{-2}$ and room temperature mobility of 147 cm$^2$/V.s is reported in MBE grown β-(Al$_{0.18}$Ga$_{0.82}$)$_2$O$_3$ / β-Ga$_2$O$_3$[11] MODFET (modulation-doped FET)[11]. Likewise, a high-density electron sheet charge of 6.4 x 10$^{12}$ cm$^{-2}$ has been reported in MOVPE-grown β-(Al$_{0.26}$Ga$_{0.74}$)$_2$O$_3$ / β-Ga$_2$O$_3$ heterostructures[12]. Based on the phase diagram of Ga$_2$O$_3$-Al$_2$O$_3$, growth of high composition (x > 0.25) β-(Al$_x$Ga$_{1-x}$)$_2$O$_3$ requires higher growth temperatures (> 800 C)[13]. On the other hand, the spread of silicon donors from delta-doped β-(Al$_x$Ga$_{1-x}$)$_2$O$_3$ layer into the UID channel layer becomes prominent at high temperatures[12], increasing ionized impurity scattering in β-Ga$_2$O$_3$ channel. Obtaining high-density 2DEG using β-(Al$_x$Ga$_{1-x}$)$_2$O$_3$ / β-Ga$_2$O$_3$ with minimal silicon spread remains an open challenge[12,14].

Recently, metastable phases of Ga$_2$O$_3$, such as ε-[15] and α-Ga$_2$O$_3$[16] are starting to garner interest because of their ultra-wide band gap and unique material properties[15]. In particular,

ε-Ga$_2$O$_3$ is an ultra-wide bandgap semiconductor and is expected to be a ferroelectric with switchable spontaneous polarization under external electric field[17]. Based on first-principles density-functional-theory (DFT) calculations, it is predicted that ε-Ga$_2$O$_3$ can be stabilized over competing α and β phases under the appropriate epitaxial strain[18]. By selecting an epitaxially-matched substrate also having a large conduction band offset with ε-Ga$_2$O$_3$, a 2DEG is expected to form at the heterointerface without any intentional doping. Because the spontaneous polarization (P$_{sp}$) of ε-Ga$_2$O$_3$ is nearly an order of magnitude higher than that of other III-V semiconductors, such as GaN and AlN[19], it is expected to lead to a high-density 2DEG at the heterointerface. Furthermore, the ferroelectric nature of ε-Ga$_2$O$_3$ with reversible P$_{sp}$ allows a more drastic change of 2DEG density via polarization switching. However, there are only a limited set of commercially available substrates that satisfy both the criteria of having good epitaxial match and large conduction band offset. Leone et.al recently reported a sheet charge density of 6.4 x 10$^{12}$ cm$^{-2}$ at ε-Ga$_2$O$_3$ / GaN heterointerface[20]. However, it is challenging to obtain high-quality heterointerfaces that could offer superior electron transport properties.

Analogous to AlGaN / GaN heterointerfaces, ε-AlGaO$_3$ / ε-Ga$_2$O$_3$ interfaces with a relatively low lattice mismatch can facilitate electron confinement. Recently, growth of ε-, κ-AlGaO$_3$ has been reported in mist-CVD[21] and PLD[15,22]. By optimizing the growth conditions, high-quality ε-AlGaO$_3$ / ε-Ga$_2$O$_3$ interfaces could potentially be obtained, enabling high-density 2DEG formation. Additionally, 2DEG at ε-AlGaO$_3$ / ε-Ga$_2$O$_3$ could lead to high mobility channel ε-Ga$_2$O$_3$, as there are no ionized donor atoms in the barrier or the channel layers. Although there are multiple reports on the polar properties of ε-Ga$_2$O$_3$[18,23,24], there is no information on polarization properties of ε-AlGaO$_3$. This unique combination of conductive ultra-wide bandgap semiconductor with a large switchable spontaneous polarization could enable applications for high-power, non-volatile memories and neuromorphic computing. To gauge the performance of ε-$Ga_2O_3$ based devices, knowledge of the electronic and polar properties of ε-$AlGaO_3$ and the confinement of electrons at the ε-$AlGaO_3$ / ε-$Ga_2O_3$ interface are needed, which is currently lacking.

In this report, we have used DFT to calculate the band gap, conduction band offsets, polarization constants, elastic and piezoelectric tensor matrices for ε-$Ga_2O_3$, κ-$Al_2O_3$ and ε-$AlGaO_3$ alloys. Using these values, we calculated the 2DEG charge density for different epitaxial barrier thickness and alloy ordering. We find that 2DEG charge densities between $3.8 \times 10^{12}$ $cm^{-2}$ and $1.4 \times 10^{14}$ $cm^{-2}$ can be attained for ε-$AlGaO_3$ / ε-$Ga_2O_3$ heterostructures. To maximize the charge contrast, we studied the effect of spontaneous polarization reversal on ε-$Ga_2O_3$ / ε-$AlGaO_3$ / ε-$Ga_2O_3$. We show that 2DEG sheet charge contrast ratios as high as 1600 can be attained for optimized ε-$Ga_2O_3$ cap and ε-$AlGaO_3$ barrier thicknesses.

For the DFT calculations, we used the VASP code[25] with projector augmented-wave[26] potentials. The plane-wave basis set was expanded to a cutoff energy of 520 eV to minimize Pulay stresses during the full relaxation, and the criteria for the relaxation was set to 0.01 eV/Å. The Brillouin zone was sampled with the Monkhorst-Pack method with grids of 6×4×4 for the ε-phase alloys. The 3d, 4s, and 4p states of Ga, 3s and 3p states of Al, 2s and 2p states of O are taken as valence states, and the exchange-correlation energy was described with the Perdew, Burke, and Ernzerhof (PBE) functional[27]. The bandgaps and electron affinities were calculated using the electrostatic potential of the non-polar (010) surface and the band gap of the bulk phase obtained using the hybrid Heyd-Scuseria-Ernzerhof (HSE) functional with a mixing parameter of 0.35[28]. We used density functional perturbation theory (DFPT) with an increased cutoff energy of 700 eV to evaluate the dielectric and piezoelectric constants[29].

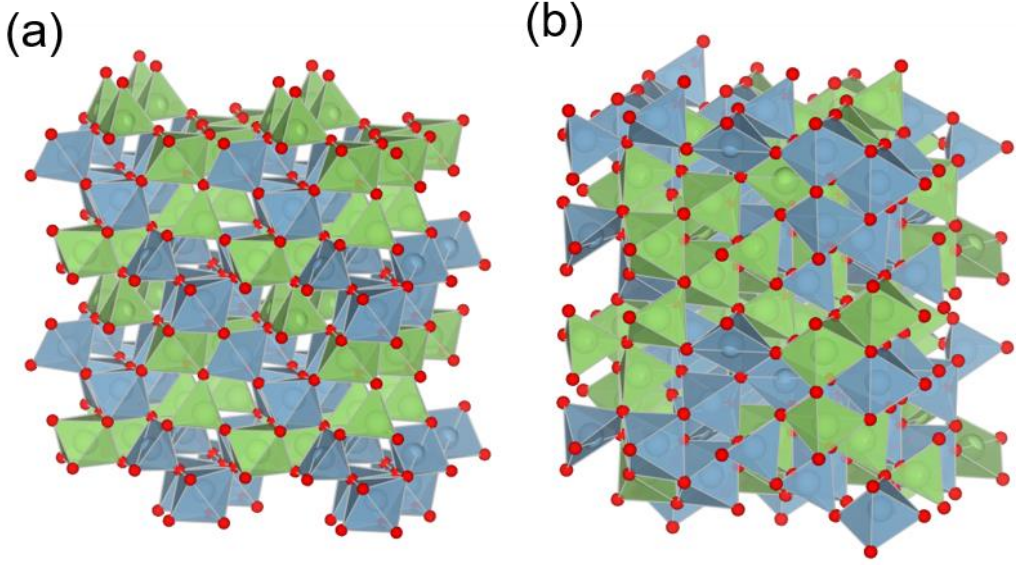

FIG. 1. Atomic models of (a) ordered- and (b) disordered AlGaO$_3$ alloys.

We considered two different structures to simulate ε-AlGaO$_3$ alloys, as shown in Figs. 1(a) and 1(b). One is an ordered alloy (Fig. 1(a)), wherein the Al/Ga cations occupy four distinguishable Wyckoff sites. The other (Fig. 1(b)) is a disordered alloy structure of ε-AlGaO$_3$ constructed using an 80-atom special quasi-random structure(SQS)[30] with perfectly disordered pair correlation. We calculated the free energy of mixing (G$_{mix}$) using the following equation:

$$G_{\text{mix}} = E(\text{Al}_{1-x}\text{Ga}_x\text{O}_3) - xE(\text{Ga}_2\text{O}_3) - (1-x)E(\text{Al}_2\text{O}_3) - TS^{vib}(\text{Al}_{1-x}\text{Ga}_x\text{O}_3) - TS^{conf}(x), \quad (1)$$

where $E(\text{Al}_{1-x}\text{Ga}_x\text{O}_3)$, $E(\text{Ga}_2\text{O}_3)$, $E(\text{Al}_2\text{O}_3)$ is the normalized total energy of SQS, Ga$_2$O$_3$, and Al$_2$O$_3$ cell, respectively, $T$ is the temperature in Kelvin, $S^{vib}$ is the vibrational entropy that can be calculated using a phonon calculation[31], and $S^{conf}$ is the configurational entropy, which in the case of the disordered alloy is $-nk_B(x \ln x + (1-x)\ln(1-x))$, where $n$ is the number of cation sites and $k_B$ is Boltzmann constant. Due to the contribution of configurational entropy in the disordered alloy, we expect an order-to-disorder transition to occur above 405 °C, as shown in Fig. 2(a). Because this transition temperature of 405 °C is less than the reported growth temperature[32] of ε-$Ga_2O_3$, which is in the range of 610 °C – 700 °C[17),20)21)], the disordered alloy is expected to be more favorable, and the ordered alloy should only be considered as a metastable phase.

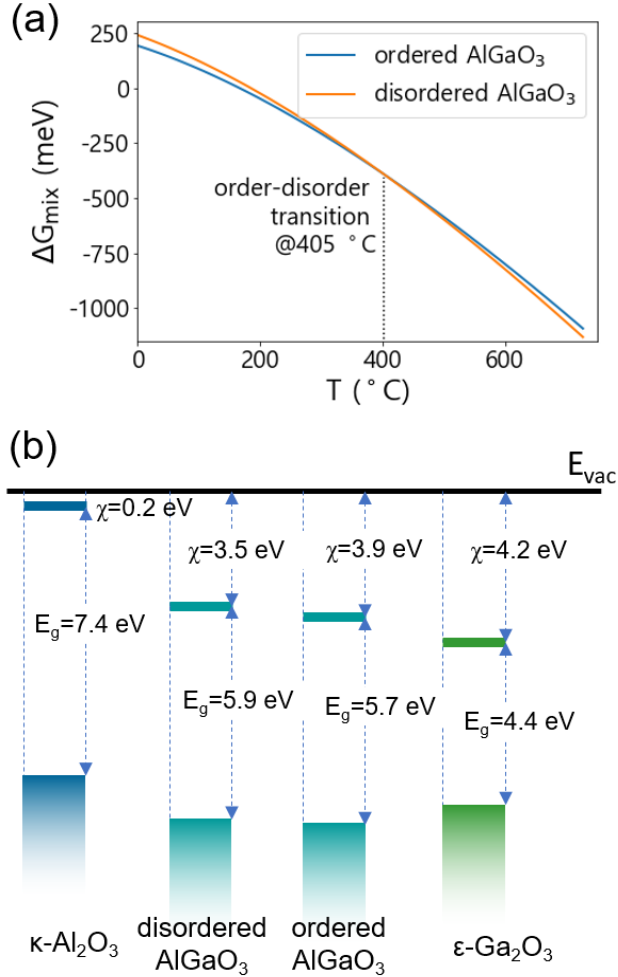

FIG. 2. (a) Gibbs free energy of mixing of ordered/disordered AlGaO₃ (b) Band alignment of κ-$Al_2O_3$, AlGaO₃, and ε-$Ga_2O_3$.

A list of key material parameters of ε-$Ga_2O_3$, κ-$Al_2O_3$ and ε-AlGaO₃ are listed in Table 1. The calculated spontaneous polarizations of ε-$Ga_2O_3$ and κ-$Al_2O_3$ are 23 μC/cm² and 26 μC/cm², respectively. The calculated polarization for ε-AlGaO$_3$ is 17.6 μC/cm² and 13 μC/cm² for the ordered and disordered phase, respectively. We note here that the polarization of the disordered structure can vary in the range of 12 – 13.6 μC/cm² depending on different SQS cells. We determined the polarization value of the disordered phase from the most stable SQS configurations. Such deviation from a linear interpolation of polarization of the end members is also reported in group III- nitride pseudobinary alloys due to hydrostatic pressure and internal strain effects[33]. The magnitude of piezoelectric tensor coefficients is also smaller in ε-AlGaO$_3$ compared to ε-Ga$_2$O$_3$. Matrix elements of the stiffness tensor are adopted to simulate the biaxial strain effect. The calculated P$_{sp}$ value of ε-Ga$_2$O$_3$ matches with other theoretical predictions[23]. The experimentally determined P$_{sp}$ of ε-Ga$_2$O$_3$ is significantly smaller (0.18 μC/cm²) than the theoretical prediction and has been attributed to the lack of high-quality thin films[17]. Currently there are no reports on theoretical or experimentally determined polarization constants for ε-AlGaO$_3$.

To form a 2DEG at the ε-AlGaO$_3$ / ε-Ga$_2$O$_3$ heterointerface, a positive bound sheet charge ($\rho_b$) is required at the junction. The sign of the bound polarization charge is dictated by the polarization discontinuity at the interface ($\nabla \cdot P = -\rho_b$). To realize a 2DEG at the ε-AlGaO$_3$ / ε-Ga$_2$O$_3$ heterointerface, the spontaneous polarization of the channel and the barrier should be aligned along the c-axis (001) (Fig. 1). Generally, when thickness of a polar material exceeds a critical value, a surface dipole forms which neutralizes the electric field in the bulk of the film[34]. On the other hand, in a heterojunction above the critical thickness, polarization sheet charge at the heterojunction induces a large band bending at the heterointerface, thereby, a 2DEG channel is formed to neutralize the positive sheet charge[35,36]. The 2DEG sheet charge density is calculated taking into account the spontaneous polarization discontinuity and piezoelectric polarization[19]. We find that the spontaneous polarization discontinuity dominates strongly over the piezoelectric polarization at ε-AlGaO$_3$ / ε-Ga$_2$O$_3$ interfaces.

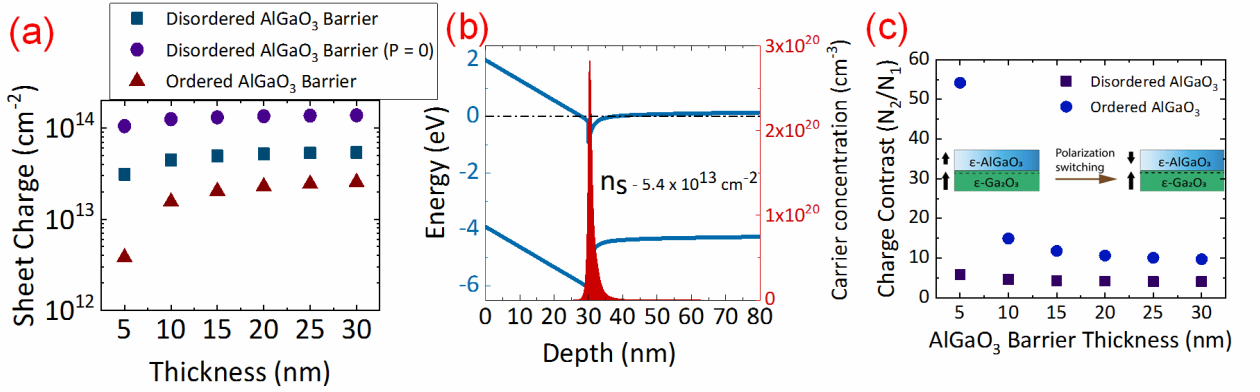

FIG. 3. (a) Calculated 2DEG sheet charge densities for various ε-AlGaO$_3$ barrier thicknesses and alloy ordering (b) Band diagram and 2DEG charge profile of disordered ε-AlGaO$_3$(30 nm) / ε-Ga$_2$O$_3$ heterojunction (c) Calculated charge contrast for ordered and disordered ε-AlGaO$_3$ barriers under polarization reversal along with the schematic of polarization switching in the heterostructure (inset)

Using a Schrödinger–Poisson solver[37)] we calculated the 2DEG sheet charge density as a function of ε-AlGaO$_3$ barrier type and barrier thickness, which was varied from 5 to 30 nm, assuming a surface barrier of 2 eV. We used an effective mass of 0.24m$_0$ and dielectric constant of 13.2ε$_0$ for both ε-AlGaO$_3$ and ε-Ga$_2$O$_3$ thin films, calculated from DFT and DFPT. Since there is no experimental data on the nature of ε-AlGaO$_3$ alloy, we considered three kinds of possible ε-AlGaO$_3$ alloy configurations, ordered ε-AlGaO$_3$, disordered ε-AlGaO$_3$ and disordered ε-AlGaO$_3$ with P = 0. As shown in Figure 3(a), the 2DEG sheet charge density increased with increasing barrier thickness from 5 nm to 30 nm. As the barrier thickness is increased, the field across ε-AlGaO$_3$ reduces until the polarization charge is completely screened by the 2DEG. For thick ε-AlGaO$_3$ barrier layers (t$_b$ ~ 30 nm) we observe that 2DEG charge density (n$_s$) approaches σ$_\pi$/e (polarization sheet charge). Because of the large spontaneous polarization of ε-Ga$_2$O$_3$, a significant amount of 2DEG charge can be attained even for very thin barrier layers in all the three cases. The ordered ε-AlGaO$_3$ barrier has the highest spontaneous polarization and smallest polarization mismatch, hence the 2DEG charge density is lowest of the three cases. In the case of the disordered ε-AlGaO$_3$ barrier with P = 0, 2DEG densities greater than 1 x 10$^{14}$ cm$^{-2}$ can be attained that are nearly independent of the thickness of the barrier layers. For the case of disordered ε-AlGaO$_3$ with a finite polarization, 2DEG sheet charge densities close to 5.4 x 10$^{13}$ cm$^{-2}$ can be obtained. The band diagram of disordered ε-AlGaO$_3$ (finite P) / ε-Ga$_2$O$_3$ heterojunction with a 30 nm barrier and 2 eV surface barrier height is shown in Fig. 3(b).

Unlike polar III-V semiconductors like GaN and AlN, ε-Ga$_2$O$_3$ is predicted to be a ferroelectric material with reversible spontaneous polarization[18]. The direction of the spontaneous polarization in ε-AlGaO$_3$ can be reversed by applying an electric field across a metal contact on the barrier layer surface and the contact to the 2DEG channel. The 2DEG sheet charge is expected to increase on reversing the polarization of ε-AlGaO$_3$ because of enhancement of polarization sheet charge at the heterointerface. To obtain a large charge and resistance contrast between the two states of polarization in the alloy barrier, the 2DEG density in state 1 (low charge state) needs to be much lower than that of state 2 (high charge state). We evaluated 2DEG charge density of ε-AlGaO$_3$ / ε-Ga$_2$O$_3$ heterojunction with ε-AlGaO$_3$ polarization along (001) and (00-1) directions. The calculated charge contrast for the two configurations is plotted in Fig. 3(c). We found that sheet charge densities can be enhanced up to 54 times using the ordered ε-AlGaO$_3$ barrier. In the case of the disordered barrier ($P_{sp}$ ~ 13 µC/cm$^2$), the charge contrast is close to 5x, since the 2DEG charge density at disordered ε-AlGaO$_3$ / ε-Ga$_2$O$_3$ interface is significant even for very thin barrier layers for both the states.

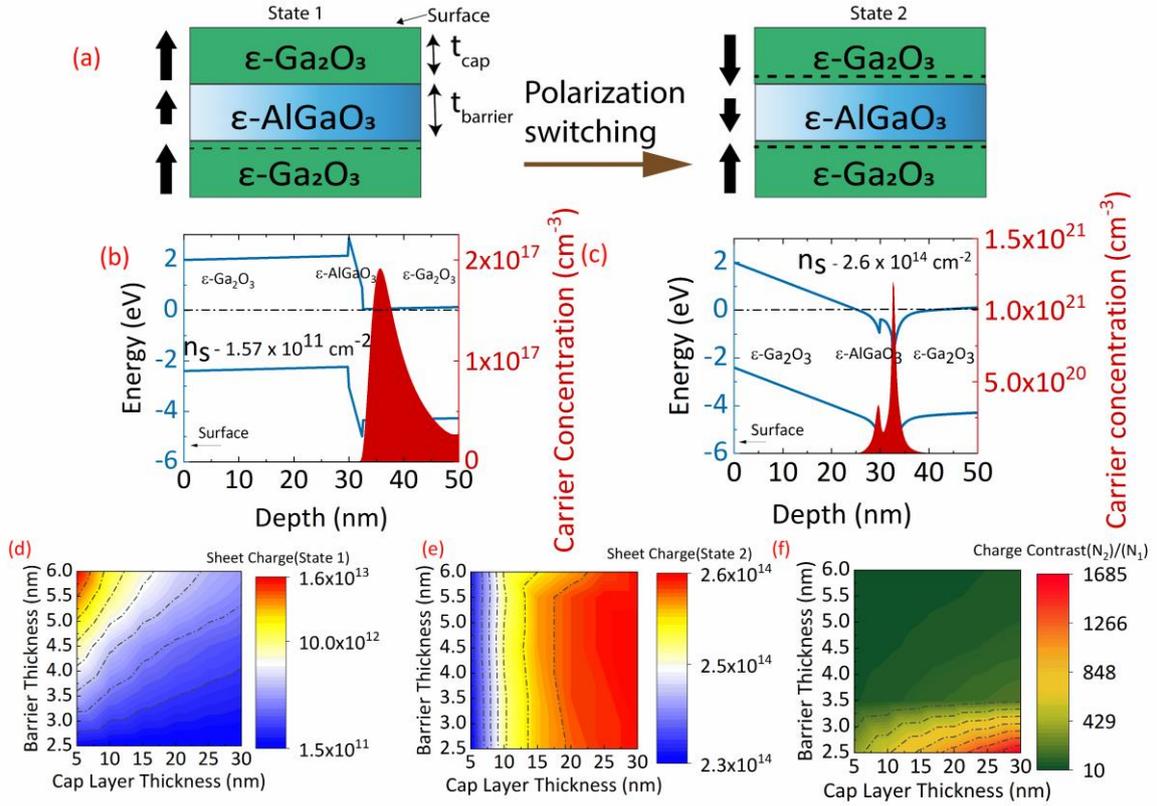

FIG. 4. (a) Schematic depicting polarization switching in $\varepsilon\text{-}Ga_2O_3 / \varepsilon\text{-}AlGaO_3 / \varepsilon\text{-}Ga_2O_3$ heterostructures. (b) Band diagram of $\varepsilon\text{-}Ga_2O_3/\varepsilon\text{-}AlGaO_3/ \varepsilon\text{-}Ga_2O_3$ heterostructure in state 1 (c) Corresponding band diagram of $\varepsilon\text{-}Ga_2O_3/\varepsilon\text{-}AlGaO_3/ \varepsilon\text{-}Ga_2O_3$ heterostructure with reversed spontaneous polarization (state 2). (d) 2D contour plot of 2DEG charge density variation with cap and barrier layer thicknesses (state 1) (e) 2D contour plot of 2DEG charge density variation with cap and barrier layer thicknesses (state 2) (f) Contour plot of charge contrast ratio between state 1 and state 2

To maximize the charge contrast for $\varepsilon\text{-}AlGaO_3 / \varepsilon\text{-}Ga_2O_3$ material system, we studied polarization switching in $\varepsilon\text{-}Ga_2O_3 /\varepsilon\text{-}AlGaO_3 / \varepsilon\text{-}Ga_2O_3$ double heterostructures. Fig. 4(a) shows the $\varepsilon\text{-}Ga_2O_3 /\varepsilon\text{-}AlGaO_3 / \varepsilon\text{-}Ga_2O_3$ double heterostructure with two different polarization configurations. In state 1, all the three layers have spontaneous polarization along the *c*-axis (001). When an electric field is applied, the polarization direction of the barrier and cap layers is reversed (state 2). This results in the formation of an additional 2DEG channel between the cap and the barrier layer. We studied the case for disordered $\varepsilon\text{-}AlGaO_3$ barrier with a finite polarization (Fig. 4 (a)). Similar analysis is also done for ordered $\varepsilon\text{-}AlGaO_3$ and disordered $\varepsilon\text{-}AlGaO_3(P = 0)$ (see supplementary data). In such a structure, the 2DEG density is a function of the thicknesses of the cap layer and the barrier.

To understand the 2DEG dependence on heterostructure design parameters, we constructed a contour map of 2DEG density as a function of ε-Ga$_2$O$_3$ and ε-AlGaO$_3$ thicknesses. Fig. 4(d) shows the 2DEG charge density dependence on design parameters on state 1. To minimize the 2DEG density in state 1($N_1$), the cap layer thickness should be maximized while the barrier thickness should be kept as small as possible. This trend is similar to the behavior of ε-AlGaO$_3$ / ε-Ga$_2$O$_3$ structure discussed earlier (Fig 3). For a thin barrier layer, the potential drop across the barrier is not enough to induce a high-density 2DEG at the ε-AlGaO$_3$ / ε-Ga$_2$O$_3$ interface. Depending on the design, 2DEG densities between 1.5 x 10$^{11}$ cm$^{-2}$ to 1.6 x 10$^{13}$ cm$^{-2}$ can be attained for state 1.

Fig. 4(b) shows the calculated band diagram and the 2DEG profile for state 1 with a cap layer thickness of 30 nm and barrier thickness of 2.5 nm. A 2DEG density of 1.5 x 10$^{11}$ cm$^{-2}$ is realized for this structure. Under polarization reversal, the 2DEG density in state 2 ($N_2$) is enhanced because of the formation of a second 2DEG channel at the ε-AlGaO$_3$ / ε-Ga$_2$O$_3$ interface. Because of a high polarization sheet charge between ε-AlGaO$_3$ / ε-Ga$_2$O$_3$ interfaces, we observe a weak dependence on design parameters. A high 2DEG sheet charge of 2.3 x 10$^{14}$ cm$^{-2}$ to 2.6 x 10$^{14}$ cm$^{-2}$ can be attained for the same parameter space (state 2). The plot of band diagram and 2DEG charge profile for state 2 is shown in Fig. 4(c). At such high 2DEG charge densities, we see poor confinement of 2DEG in ε-Ga$_2$O$_3$ channel layers. Hence, we see a significant amount of 2DEG electron charge in ε-AlGaO$_3$ barrier layer (Fig. 4(e)). The charge contrast ratio ($N_2$ / $N_1$) between the two states is plotted in Fig. 4(f). As explained before, the charge ratio is maximum for designs with a thin barrier layer and thick cap layer. Charge contrast ratio as high as 1600 can be achieved using ε-Ga$_2$O$_3$ /ε-AlGaO$_3$ / ε-Ga$_2$O$_3$ double heterostructure.

Although this report of 2DEG formation at ε-AlGaO$_3$ / ε-Ga$_2$O$_3$ interface is encouraging, heteroepitaxial ferroelectric ε-Ga$_2$O$_3$ films suffer from issues such as domain formation[38], structural defects and formation of mixed phases[39,40]. Significant experimental work[41] needs to be done to understand the epitaxy of high-quality, phase pure ε-Ga$_2$O$_3$ [42] and ε-AlGaO$_3$. Also, detailed polarization studies of ε-Ga$_2$O$_3$ and its alloys are needed to understand 2DEG formation at ε-AlGaO$_3$ / ε-Ga$_2$O$_3$ interface. Nonetheless, we expect this computational report of polarization reversal in ε-Ga$_2$O$_3$ /ε-AlGaO$_3$ / ε-Ga$_2$O$_3$ heterostructures for obtaining high charge contrast in the ε-Ga$_2$O$_3$ material system, will serve as a motivation for further experimental studies.

In summary, we have calculated relevant material properties of ε-Ga$_2$O$_3$ and ε-AlGaO$_3$ alloys using DFT. Utilizing these values, we calculated 2DEG charge densities for ordered ε-AlGaO$_3$, disordered ε-AlGaO$_3$, and disordered ε-AlGaO$_3$ (P = 0). 2DEG charge densities between $3.8 \times 10^{12}$ cm$^{-2}$ to $1.4 \times 10^{14}$ cm$^{-2}$ can be obtained depending on barrier type and ε-AlGaO$_3$ thickness. To maximize the charge contrast between state 1 and state 2, we studied spontaneous polarization switching in ε-Ga$_2$O$_3$ / ε-AlGaO$_3$ / ε-Ga$_2$O$_3$ heterostructures. Charge contrast ratios as high as 1600 can be obtained using a thick ε-Ga$_2$O$_3$ cap layer and thin ε-AlGaO$_3$ layers. This report shows the high potential of ε-AlGaO$_3$ / ε-Ga$_2$O$_3$ heterostructures for non-volatile memory and neuromorphic applications.

**Acknowledgements**

We acknowledge funding from the National Science Foundation (NSF) through grants DMR-1931610 and DMR-1931652. This material is also based upon work supported by the Air Force Office of Scientific Research under award number FA9550-18-1-0507. Any opinions, finding, and conclusions or recommendations expressed in this material are those of the author and do not necessarily reflect the views of the United States Air Force. We also gratefully acknowledge partial support from the National Research Foundation of Korea (NRF-2019R1F1A1058554). The computations were carried out using the Extreme Science and Engineering Discovery Environment (XSEDE), which is supported by NSF ACI-1548562, and resources from Korea Supercomputing Center (KSC-2019-CRE-0023).

**Figure Captions**

**Fig. 1.** Atomic models of (a) ordered- and (b) disordered AlGaO$_3$ alloys.

**Fig. 2.** (a) Gibbs free energy of mixing of ordered/disordered AlGaO$_3$ (b) Band alignment of κ-Al$_2$O$_3$, AlGaO$_3$, and ε-Ga$_2$O$_3$.

**Fig. 3.** (a) Calculated 2DEG sheet charge densities for various ε-AlGaO$_3$ barrier thicknesses and alloy ordering (b) Band diagram and 2DEG charge profile of disordered ε-AlGaO$_3$(30 nm) / ε-Ga$_2$O$_3$ heterojunction (c) Calculated charge contrast for ordered and disordered ε-AlGaO$_3$ barriers under polarization reversal along with the schematic of polarization switching in the heterostructure (inset).

**Fig. 4.** (a) Schematic depicting polarization switching in ε-Ga$_2$O$_3$ / ε-AlGaO$_3$ / ε-Ga$_2$O$_3$ heterostructures. (b) Band diagram of ε-Ga$_2$O$_3$/ε-AlGaO$_3$/ ε-Ga$_2$O$_3$ heterostructure in state 1 (c) Corresponding band diagram of ε-Ga$_2$O$_3$/ε-AlGaO$_3$/ ε-Ga$_2$O$_3$ heterostructure with reversed spontaneous polarization (state 2). (d) 2D contour plot of 2DEG charge density variation with cap and barrier layer thicknesses (state 1) (e) 2D contour plot of 2DEG charge density variation with cap and barrier layer thicknesses (state 2) (f) Contour plot of charge contrast ratio between state 1 and state 2 (a) Schematic depicting polarization switching in ε-Ga$_2$O$_3$ / ε-AlGaO$_3$ / ε-Ga$_2$O$_3$ heterostructures. (b) Band diagram of ε-Ga$_2$O$_3$/ε-AlGaO$_3$/ ε-Ga$_2$O$_3$ heterostructure in state 1 (c) Corresponding band diagram of ε-Ga$_2$O$_3$/ε-AlGaO$_3$/ ε-Ga$_2$O$_3$ heterostructure with reversed spontaneous polarization (state 2). (d) 2D contour plot of 2DEG charge density variation with cap and barrier layer thicknesses (state 1) (e) 2D contour plot of 2DEG charge density variation with cap and barrier layer thicknesses (state 2) (f) Contour plot of charge contrast ratio between state 1 and state 2.

**Table I.** DFT calculated material parameters for ε-Ga$_2$O$_3$, κ-Al$_2$O$_3$ and ε-AlGaO$_3$ alloys DFT calculated material parameters for ε-Ga$_2$O$_3$, κ-Al$_2$O$_3$ and ε-AlGaO$_3$ alloys

| | P$_{sp}$ (μC/cm$^2$) | e$_{31}$ (μC/cm$^2$) | e$_{32}$ (μC/cm$^2$) | e$_{33}$ (μC/cm$^2$) | C$_{13}$ (GPa) | C$_{23}$ (GPa) | C$_{33}$ (GPa) |
|---|---|---|---|---|---|---|---|
| ε-Ga$_2$O$_3$ | 23 | 9.5 | 7.9 | -16.3 | 125 | 125 | 207 |
| ε-AlGaO3 (ordered) | 17.6 | 6.6 | 2.8 | -10.2 | 100 | 120 | 300 |
| ε-AlGaO3 (disordered) | 13 | 7.2 | 5.7 | -11.2 | 137 | 120.5 | 291.5 |
| κ-Al$_2$O$_3$ | 26 | 4.6 | 3.5 | -5.6 | 149 | 116 | 376 |

**FIGURES**

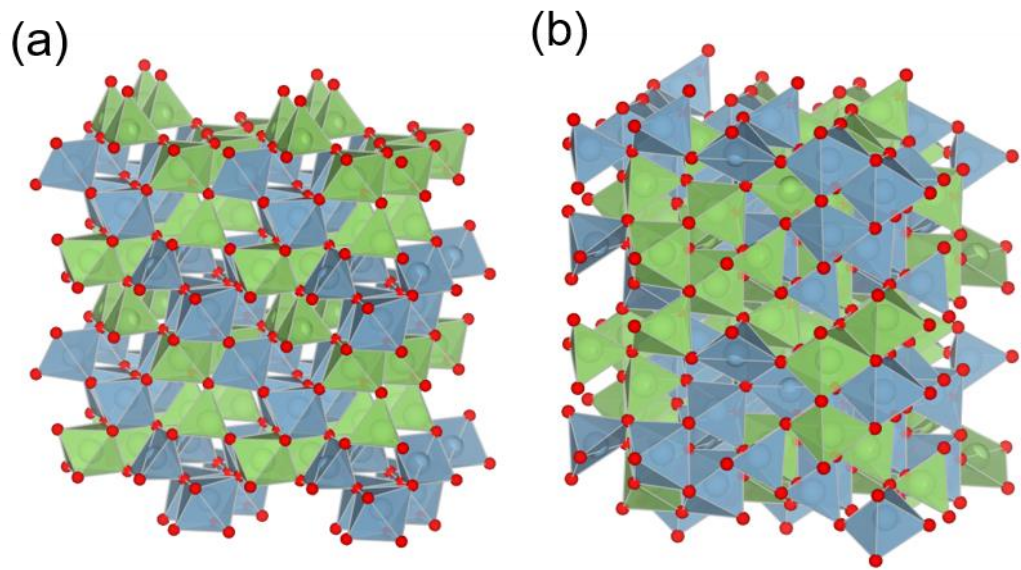

FIG. 1. Atomic models of (a) ordered- and (b) disordered AlGaO$_3$ alloys.

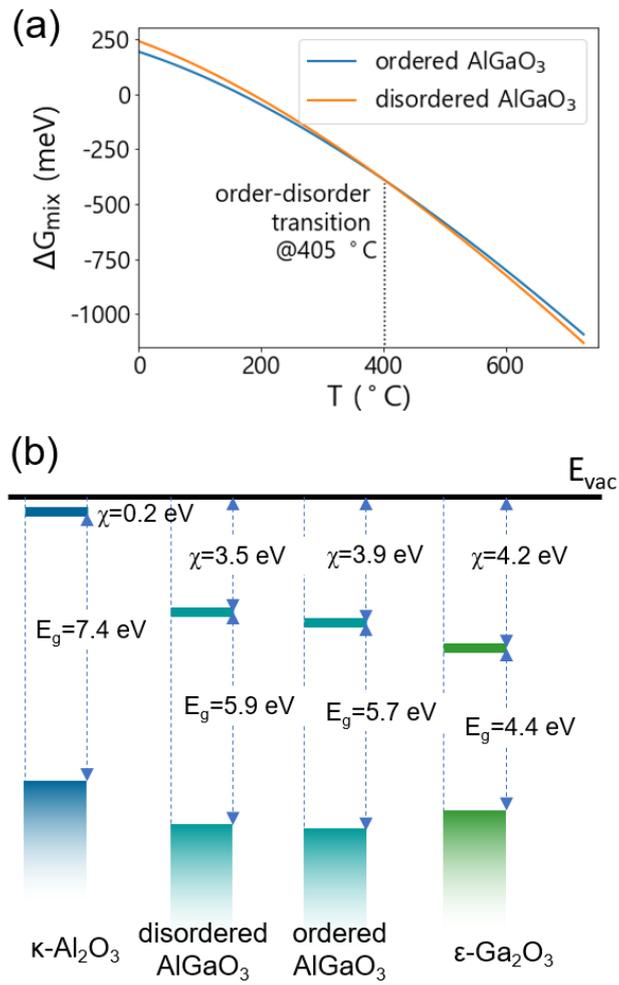

FIG. 2. (a) Gibbs free energy of mixing of ordered/disordered AlGaO$_3$ (b) Band alignment of κ-Al$_2$O$_3$, AlGaO$_3$, and ε-Ga$_2$O$_3$.

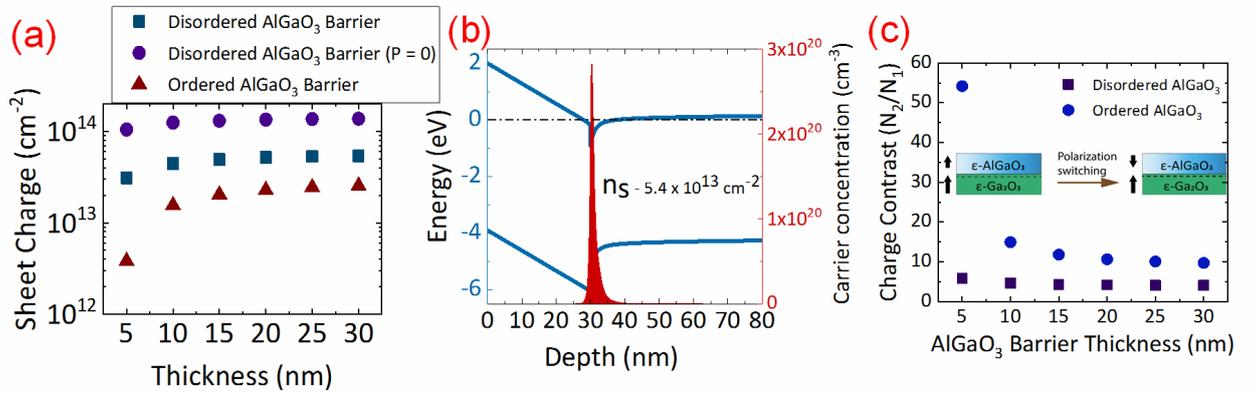

FIG. 3. (a) Calculated 2DEG sheet charge densities for various ε-AlGaO$_3$ barrier thicknesses and alloy ordering (b) Band diagram and 2DEG charge profile of disordered ε-AlGaO$_3$(30 nm) / ε-Ga$_2$O$_3$ heterojunction (c) Calculated charge contrast for ordered and disordered ε-AlGaO$_3$ barriers under polarization reversal along with the schematic of polarization switching in the heterostructure (inset).

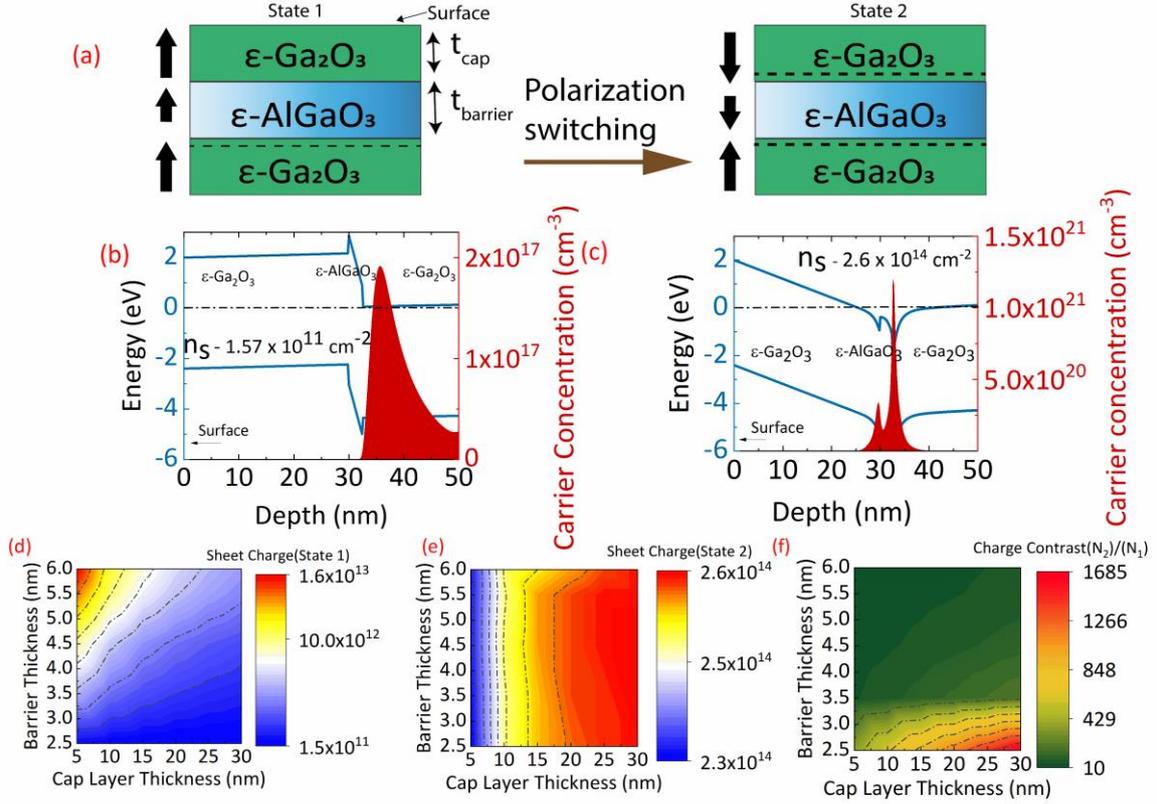

FIG. 4. (a) Schematic depicting polarization switching in ε-Ga$_2$O$_3$ / ε-AlGaO$_3$ / ε-Ga$_2$O$_3$ heterostructures. (b) Band diagram of ε-Ga$_2$O$_3$/ε-AlGaO$_3$/ ε-Ga$_2$O$_3$ heterostructure in state 1 (c) Corresponding band diagram of ε-Ga$_2$O$_3$/ε-AlGaO$_3$/ ε-Ga$_2$O$_3$ heterostructure with reversed spontaneous polarization (state 2). (d) 2D contour plot of 2DEG charge density variation with cap and barrier layer thicknesses (state 1) (e) 2D contour plot of 2DEG charge density variation with cap and barrier layer thicknesses (state 2) (f) Contour plot of charge contrast ratio between state 1 and state 2.

**Supplementary information**

**Highly tunable, polarization-engineered two-dimensional electron gas in ε-AlGaO$_3$ / ε-Ga$_2$O$_3$ heterostructures**

*Praneeth Ranga,[1*,] Sung Beom Cho[2,3*], Rohan Mishra[2] and Sriram Krishnamoorthy[1]*

[1] *Department of Electrical and Computer Engineering, The University of Utah, Salt Lake City, UT 84112, United States of America*

[2] *Department of Mechanical Engineering and Materials Science, and Institute of Materials Science and Engineering, Washington University in St. Louis, St. Louis, Missouri 63130, USA*

[3] *Virtual Engineering Center, Korea Institute of Ceramic Engineering and Technology, Jinju-si, Republic of Korea*

\* indicates equal contribution

Email:praneeth.ranga@utah.edu,rmishra@wustl.edu,sriram.krishnamoorthy@utah.edu

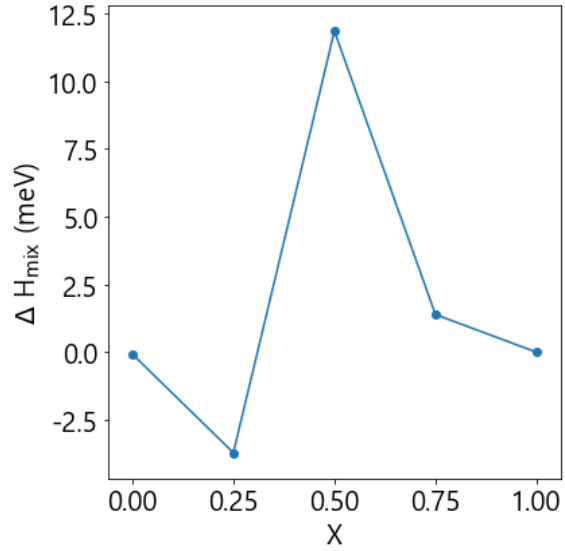

**Fig. S1** Enthalpy of mixing between *Pna2₁* Al$_2$O$_3$-Ga$_2$O$_3$ alloy

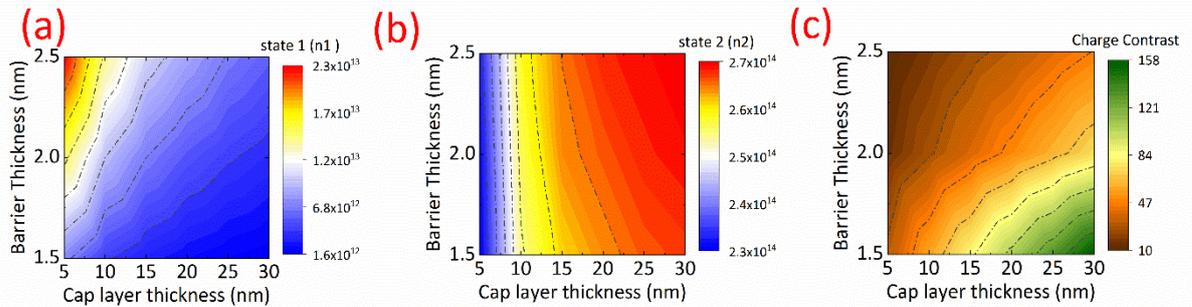

**Fig. S2** **(a)** Contour plot of 2DEG density with cap and barrier layer thickness of disordered ε-AlGaO (P = 0) (state 1) (b) Contour plot of 2DEG change with cap and barrier layer thickness with disordered ε-AlGaO (P = 0) (state 2) (c) Contour plot of 2DEG change charge contrast between state 1 and state 2

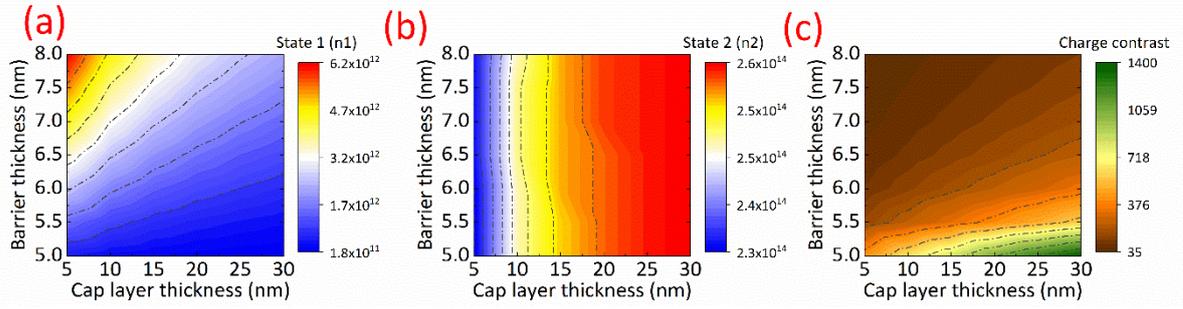

**Fig. S3 (a)** Contour plot of 2DEG density with cap and barrier layer thickness of ordered ε-AlGaO (state 1) (b) Contour plot of 2DEG change with cap and barrier layer thickness with ordered ε-AlGaO (state 2) (c) Contour plot of 2DEG change charge contrast between state 1 and state 2

**Table I** Calculated stiffness tensor of ordered ε-AlGaO$_3$ (GPa)

|     | xx  | yy  | zz  | xy  | yz  | zx  |
|-----|-----|-----|-----|-----|-----|-----|
| xx  | 310 | 134 | 100 |     |     |     |
| yy  | 134 | 311 | 120 |     |     |     |
| zz  | 100 | 120 | 300 |     |     |     |
| xy  |     |     |     | 115 |     |     |
| yz  |     |     |     |     | 78  |     |
| zx  |     |     |     |     |     | 48  |

**Table II** Calculated stiffness tensor of disordered ε-AlGaO$_3$ (GPa)

|     | xx    | yy    | zz    | xy   | yz   | zx   |
|-----|-------|-------|-------|------|------|------|
| xx  | 275   | 131.5 | 137   |      |      |      |
| yy  | 131.5 | 267.5 | 120.5 |      |      |      |
| zz  | 137   | 120.5 | 291.5 |      |      |      |
| xy  |       |       |       | 65.5 |      |      |
| yz  |       |       |       |      | 76.5 |      |
| zx  |       |       |       |      |      | 48.5 |

**Table III** Calculated stiffness tensor of ε-Ga$_2$O$_3$ (GPa)

|     | xx  | yy  | zz  | xy | yz | zx |
|-----|-----|-----|-----|----|----|----|
| xx  | 217 | 144 | 125 |    |    |    |
| yy  | 144 | 192 | 125 |    |    |    |
| zz  | 125 | 125 | 207 |    |    |    |
| xy  |     |     |     | 55 |    |    |
| yz  |     |     |     |    | 29 |    |
| zx  |     |     |     |    |    | −1 |

**Table IV** Calculated stiffness tensor of κ-Al$_2$O$_3$ (GPa)

|     | xx  | yy  | zz  | xy | yz  | zx |
|-----|-----|-----|-----|----|-----|-----|
| xx  | 333 | 119 | 149 |    |     |     |
| yy  | 119 | 343 | 116 |    |     |     |
| zz  | 149 | 116 | 376 |    |     |     |
| xy  |     |     |     | 76 |     |     |
| yz  |     |     |     |    | 124 |     |
| zx  |     |     |     |    |     | 98  |

**Table V.**
Calculated piezoelectric tensor of ε-Ga$_2$O$_3$ (μC/cm$^2$)

|   | xx | yy | zz | xy | yz | zx |
|---|----|----|----|----|----|----|
| x |    |    |    |    |    | 9.3 |
| y |    |    |    |    | 6.7 |    |
| z | 9.5 | 7.9 | −16.3 |    |    |    |

**Table VI.**
Calculated piezoelectric tensor of Al$_2$O$_3$ (μC/cm$^2$)

|   | xx | yy | zz | xy | yz | zx |
|---|----|----|----|----|----|----|
| x |    |    |    |    |    | 5.2 |
| y |    |    |    |    | 2.6 |    |
| z | 4.6 | 3.5 | −5.6 |    |    |    |

**Table VII.**
Calculated piezoelectric tensor of ordered ε-AlGaO$_3$ (μC/cm$^2$)

|   | xx | yy | zz | xy | yz | zx |
|---|----|----|----|----|----|----|
| x |    |    |    |    |    | 9.0 |
| y |    |    |    |    | 4.8 |    |
| z | 6.6 | 2.8 | −10.2 |    |    |    |

**Table VIII.**
Calculated piezoelectric tensor of disordered ε-AlGaO$_3$ (μC/cm$^2$)

|   | xx  | yy  | zz    | xy | yz  | zx  |
|---|-----|-----|-------|----|-----|-----|
| x |     |     |       |    |     | 7.6 |
| y |     | 2   | 5     |    | 4.8 | 0.1 |
| z | 7.2 | 5.7 | −11.2 |    |     |     |

**Table IX.**

Calculated lattice constants for κ-Al$_2$O$_3$, ordered ε-AlGaO$_3$, disordered ε-AlGaO$_3$ and ε-Ga$_2$O$_3$

|   | κ-Al$_2$O$_3$ | AlGaO$_3$ (ordered) | AlGaO$_3$ (disordered) | ε-Ga$_2$O$_3$ |
|---|---------------|---------------------|------------------------|---------------|
| A | 4.886         | 4.985               | 5.056                  | 5.127         |
| B | 8.397         | 8.564               | 8.601                  | 8.808         |
| C | 9.025         | 9.252               | 9.232                  | 9.424         |